\documentclass[preprint,prd,floatfix,nofootinbib,superscriptaddress,showpacs]{revtex4}
 \usepackage{graphicx}
\def\Z0{${\em Z^0\/}$}

\def\r#1 {$^{#1}$}

\newcommand{\gevc} { {\rm GeV/c}}
\newcommand{\gevcc}{ {\rm GeV/c^2}}

\def\gepsfcentered#1{
  \def\testit{#1}
  \def\lbracket{[}
  \ifx\testit\lbracket
    \let\dofilecmd=\gepsfwithopt
  \else
    \let\dofilecmd=\gepsfnoopt
  \fi
  \dofilecmd}

\def\gepsfnoopt#1{
  \begin{center}
  \leavevmode
  \epsffile{#1}
  \end{center}}

\def\gepsfwithopt#1 #2 #3 #4]#5{
  \begin{center}
  \leavevmode
  \gepsfmaxx=0.94\textwidth
  \epsffile[#1 #2 #3 #4]{#5}
  \end{center}}

\newdimen\gepsfmaxx
\def\epsfsize#1#2{
  \ifnum \epsfxsize=0
    \ifnum \epsfysize=0
      \ifnum #1 > \gepsfmaxx
        \gepsfmaxx
      \else
        #1
      \fi
    \else
      \epsfxsize
    \fi
  \else
    \epsfxsize
  \fi
}

\begin{document}
 \bibliographystyle{apsrev}
 \title{Search for narrow resonances below the $\Upsilon$ mesons}
 \vskip 0.025in
 \affiliation{Laboratori Nazionali di Frascati, Istituto Nazionale 
              di Fisica Nucleare, Frascati, Italy\\}
 \affiliation{Fermi National Accelerator Laboratory, Batavia, 
              Illinois 60510, USA\\}
 \affiliation{Harvard University, Cambridge, Massachusetts 02138, USA\\}
 \affiliation{Ernest Orlando Lawrence Berkeley National Laboratory, 
              Berkeley, California 94720, USA\\}
 \affiliation{Istituto Nazionale di Fisica Nucleare, University and
              Scuola Normale Superiore of Pisa, I-56100 Pisa, Italy\\}
 \affiliation{Universita di Padova, Istituto Nazionale di Fisica Nucleare,
	      Sezione di Padoova, I-35131, Padova, Italy\\}
 \vspace{0.2em}
 \author{G.~Apollinari}
 \affiliation{Fermi National Accelerator Laboratory, Batavia, 
              Illinois 60510, USA\\}
 \author{M.~Barone}
 \affiliation{Laboratori Nazionali di Frascati, Istituto Nazionale 
              di Fisica Nucleare, Frascati, Italy\\}
 \author{W.~Carithers}
 \affiliation{Ernest Orlando Lawrence Berkeley National Laboratory, 
              Berkeley, California 94720, USA\\}
 \author{M.~Dell'Orso}
 \affiliation{Istituto Nazionale di Fisica Nucleare, University and
              Scuola Normale Superiore of Pisa, I-56100 Pisa, Italy\\}
 \author{T.~Dorigo}
 \affiliation{Universita di Padova, Istituto Nazionale di Fisica Nucleare,
	      Sezione di Padoova, I-35131, Padova, Italy\\}
 \author{I.~Fiori}
 \affiliation{Istituto Nazionale di Fisica Nucleare, University and
              Scuola Normale Superiore of Pisa, I-56100 Pisa, Italy\\}
 \author{M.~Franklin}
 \affiliation{Harvard University, Cambridge, Massachusetts 02138, USA\\}
 \author{P.~Giannetti}
 \affiliation{Istituto Nazionale di Fisica Nucleare, University and
              Scuola Normale Superiore of Pisa, I-56100 Pisa, Italy\\}
 \author{P.~Giromini}
 \author{F.~Happacher}
 \author{S.~Miscetti}
 \author{A.~Parri}
 \author{F.~Ptohos}
 \altaffiliation[Present address:]{University of Cyprus, 1678 Nicosia, 
                 Cyprus\\}
 \affiliation{Laboratori Nazionali di Frascati, Istituto Nazionale 
              di Fisica Nucleare, Frascati, Italy\\}
 \author{G.~Velev}
 \affiliation{Fermi National Accelerator Laboratory, Batavia, 
              Illinois 60510, USA\\}
 \vspace{0.2em}
 \begin{abstract}
   We have investigated the invariant mass spectrum of dimuons collected
   by the CDF experiment during the $1992-1995$ run of the Fermilab Tevatron
   collider to improve the limit on the existence of narrow resonances
   set by the experiments at the SPEAR $e^{+} e^{-}$ collider. 
   In the mass range $6.3-9.0 \; \gevcc$, we derive 90\% upper credible
   limits to the ratio of the production cross section times muonic branching
   fraction of possible narrow resonances to that of the
   $ \Upsilon(1S)$ meson. In this mass range, the average limit varies from
   1.7 to 0.5\%. This limit is much worse at the mass of 7.2 $\gevcc$ due 
   to an excess of $250 \pm 61$ events with a width consistent with the
   detector resolution. \\
 \end{abstract}

 \preprint{FERMILAB-PUB-05-272-E}

 \maketitle

 \section {Introduction} \label{sec:ss-intro}
   Supersymmetric theories provide a mechanism to break the electroweak
   symmetry and to stabilize the large hierarchy between the Planck and
   the Fermi scales. Supersymmetry requires the existence of scalar 
   partners to each standard model (SM) fermion, and spin$-1/2$ partners
   to the gauge and Higgs bosons. In particular, supersymmetry predicts
   the existence of scalar quarks, i.e. particles that carry color, but 
   no spin. Scalar quarks (squarks) have been searched for at current 
   and past colliders, but charge$-1/3$ squarks might have been overlooked
   for several reasons. Within the minimal supersymmetric standard model
   (MSSM), Ref.~\cite{carena} shows that the existence of a yet undetected
   charge$-1/3$ scalar quark, lighter than the $b$ quark, would require a
   lot of fine-tuning of the MSSM parameters, but at present cannot be 
   ruled out by the electroweak precision data and the Higgs mass constraints
   from LEP2. Charge$-1/3$ squarks would give a 2\% contribution to the
   inclusive cross section for $e^{+} e^{-} \rightarrow $ hadrons; this
   contribution is comparable to the experimental error of the present
   measurements~\cite{pdg,janot}. Searches for narrow resonances at SPEAR
   have set upper bounds on $\Gamma_l$, the leptonic width of possible 
   resonances, of $100$ eV in the mass region $5.7\leq E_{\rm cm}\leq 6.4$ GeV
   and of approximately $60$ eV in the region $7.0\leq E_{\rm cm}\leq 7.4$ GeV
   ~\cite{spear}. In Ref.~\cite{nappi}, the leptonic width of 1$^{--}$ bound
   states of such squarks has been evaluated using potential models of
   ordinary heavy quarks~\cite{pot-mod}. Because of the p-wave suppression
   of the fermion contribution to their decay width, the leptonic width is
   estimated to be approximately $18$ ($6$) eV for a resonance with a $6$ 
   ($10$) $\gevcc$ mass. As noted in Ref.~\cite{nappi}, for quarkonium masses
   above $6\; \gevcc$ the width $\Gamma_l$ is well below the experimental
   bounds, and scalar-quark resonances might have been missed. With this 
   study, we investigate the region above $6.3\; \gevcc$ by using muon pairs
   with invariant mass between the $\psi$ and $\Upsilon$ mesons. The large
   statistics data set has been collected with the Collider Detector at
   Fermilab (CDF) during the $1992-1995$ Fermilab collider run. 
   Section~\ref{det} describes the detector systems relevant to this analysis.
   Section~\ref{sec:onia} reviews the expectation for bound states of charge
   $-1/3$ squarks. The data sample is described in Sec.~\ref{sec:ss-mudat}, 
   and Section~\ref{sec:ss-muan} presents additional selection criteria, 
   tuned with $\Upsilon(1S)$ decays, which reduce the non-resonant background
   by a factor of three without losing more than 10\% of the signal.
   In Sec.~\ref{sec:ss-mueps}, we fit the dimuon invariant mass distribution  
   and derive a $90$\% Bayesian upper limit on $\Gamma_l$ as a function of 
   the resonance mass. The shape of the invariant mass distribution is 
   generally quite smooth and we improve the SPEAR limit by an order of 
   magnitude. An exception is the mass of $7.2\; \gevcc$, at which the data 
   can accommodate a Gaussian bump of $250 \pm 61$ events. 
   In Sec.~\ref{sec:ss-mucheck} we explore the possibility of observing
   a real signal, we estimate its statistical significance, and we study its
   robustness by using a number of different kinematical selections.
   Our conclusions are summarized in Sec.~\ref{sec:concl}.
 \section{CDF detector} \label{det}
   CDF is a multipurpose detector, equipped with a charged particle
   spectrometer and a finely segmented calorimeter. In this section, we 
   recall the detector components that are relevant to this analysis. 
   The description of these subsystems can be found in Ref~\cite{cdfdet}.
   Two devices inside the $1.4$ T solenoid are used for measuring the
   momentum of charged particles: the silicon vertex detector (SVX) and
   the central tracking chambers (CTC). The SVX consists of four concentric
   layers of silicon microstrip detectors surrounding the beam pipe. The 
   CTC is a cylindrical drift chamber containing $84$ sense wire layers
   grouped into nine alternating superlayers of axial and stereo wires.
   Electromagnetic (CEM) and hadronic (CHA) calorimeters surround the 
   tracking volume and measure energy deposits over the pseudorapidity 
   region $|\eta| \leq$ 1. Muons are reconstructed by matching track
   segments in the drift chamber systems located outside the CHA
   (CMU, CMP, and CMX muon detectors, which cover the region $|\eta|\leq 1$)
   to the tracks of charged particles reconstructed in the CTC. The dimuon 
   events used in this analysis were collected with a three-level trigger 
   system. The first level required two charged tracks in the muon chambers.
   The second level trigger required that both muon tracks match a charged
   particle with transverse momentum $p_T \geq 2.2\; \gevc$ as
   measured by a fast track processor (CFT). The third level software 
   trigger requires that two charged CTC tracks, fully reconstructed
   in three dimensions, match track segments in the muon chambers and that
   the dimuon invariant mass is larger than $2.8\; \gevcc$.
 \section{Search for narrow resonances} \label{sec:onia}
   In this study, we search for narrow resonances {\Large $\varepsilon$},
   bound states of scalar quarks, in the dimuon invariant mass distribution
   between  $6.3$ and $9.0\; \gevcc$. For a charge$-1/3$ squark, the 
   muonic width, 
   $\Gamma_{\mu}$({\Large $\varepsilon$} $\rightarrow \mu^{+}\mu^{-})$,
   of 2P resonances in this mass range has been evaluated in Ref.~\cite{nappi}
   to be approximately $15$ eV. In contrast, the $\Upsilon(1S)$ meson has a
   larger leptonic width $\Gamma_{\mu}$ of $1.32$ keV. In analogy with the
   $\Upsilon(1S)$ meson, the annihilation of a {\Large $\varepsilon$} state
   into hadrons is believed to proceed through gluons~\cite{nappi}, the
   dominant contribution coming from the minimum number of intermediate gluons
   (one-gluon is excluded by color conservation and two-gluons by
   C-conservation). In $p\bar{p}$ collisions quarkonia states are directly
   produced through subprocesses such as  $gg \rightarrow g \Upsilon(1S)$ or
   $gg \rightarrow g \mbox{ {\Large  $\varepsilon$}}$. As the production
   subprocesses are directly related by crossing to the corresponding decay 
   processes, the production cross section $\sigma$ is determined by the 
   decay widths~\cite{coll-phys}, and approximately reads:
 \small
 \begin{equation}
 \sigma_{\mbox{\Large $\varepsilon$}}
         \; B(\mbox{{\Large $\varepsilon$}} \rightarrow \mu\mu) \simeq
     \left( \frac {m_{\Upsilon(1S)}} {m_{\mbox{\Large $\varepsilon$ }} }
     \right)^{3}
     \frac { \Gamma_{\mu }^{\mbox{\Large $\varepsilon$}} }
           { \Gamma_{\mu }^{\Upsilon(1S)} } \;
           \sigma_{\Upsilon(1S)}\;  B(\Upsilon(1S) \rightarrow \mu \mu) =
      R \; \sigma_{\Upsilon(1S)}\;  B(\Upsilon(1S) \rightarrow \mu \mu)
 \end{equation}
 \normalsize
   where $B$ is the branching ratio of the muonic decay and
   $R=\left( \frac {m_{\Upsilon(1S)}} {m_{\mbox{\Large $\varepsilon$ }} }
      \right)^{3} \; \frac { \Gamma_{\mu }^{\mbox{\Large $\varepsilon$}} }
           { \Gamma_{\mu }^{\Upsilon(1S)} }$.
   For an {\Large $ \varepsilon$} particle in the mass region investigated 
   by this study, $R$ is approximately $2$\% when the width 
   $\Gamma_{\mu}^{\mbox{\Large $\varepsilon$}}$ is evaluated using the
   standard potential of heavy-quark spectroscopy~\cite{nappi}. Using a
   different potential model and one-loop corrections to the static potential
   of the scalar quark-anti-quark system, Reference~\cite{moxhay} predicts 
   leptonic widths that are a factor of three larger. In conclusion, $R$ is
   expected to be between $2$ and $6$\%. Since CDF has collected approximately
   $10^{4}$ $\Upsilon(1S)$ mesons, the data could contain at least $200$ 
   events contributed by a hypothetical {\Large $\varepsilon$} meson
   on top of the smooth background due to Drell-Yan production, double
   semileptonic decays of $c\bar{c}$ and $b\bar{b}$ pairs, and fake muons
   produced by hadrons that mimic their signal.
 \section{Dimuon data sample} \label{sec:ss-mudat} 
   The dimuon sample used in this analysis corresponds to approximately 
   $110$ pb$^{-1}$ of data collected with the CDF detector during the
   1992$-$1995 collider run. This data set has been used in several
   CDF analyses and is described in more detail in Ref.~\cite{dimuan}.
   The muon identification is based on the three-dimensional matching of
   the track segment in the muon chambers with the track reconstructed in
   the CTC and on the energy deposited in the calorimeter tower in the muon
   path~\cite{dimuan,wenzel}. In this study, we search for muons with 
   $p_T \geq 3\; \gevc$ using the same selection criteria
   of Ref.~\cite{wenzel}; we select muons with
   $2\leq p_T \leq 3\; \gevc$ using the stricter cuts of the SLT
   algorithm~\cite{cdf-evidence,kestenbaum} in order to reduce the 
   misidentification background. We require that at least one of the muons
   is identified by both the CMP and CMU systems ($p_T \geq 3 \; \gevc$
   and pseudorapidity $|\eta| \leq 0.6$). Additional muons are
   identified by either CMU or CMX system ($p_T \geq 2 \; \gevc$
   and  $|\eta| \leq 1.0$). 
   These selection criteria yield reconstructed muon pairs with rapidity
   $|y| \leq 1.0$. We retain events which contain two and only two muons.
   The muon momentum is evaluated with a fit which constrains the track 
   to originate from the beam line. The dimuon invariant mass is calculated
   using these momenta. This study uses opposite charge dimuon pairs; 
   however, distributions for same charge dimuon pairs are also shown as a
   cross-check.

\begin{figure}[t]
 \begin{center}
 \leavevmode
 \includegraphics{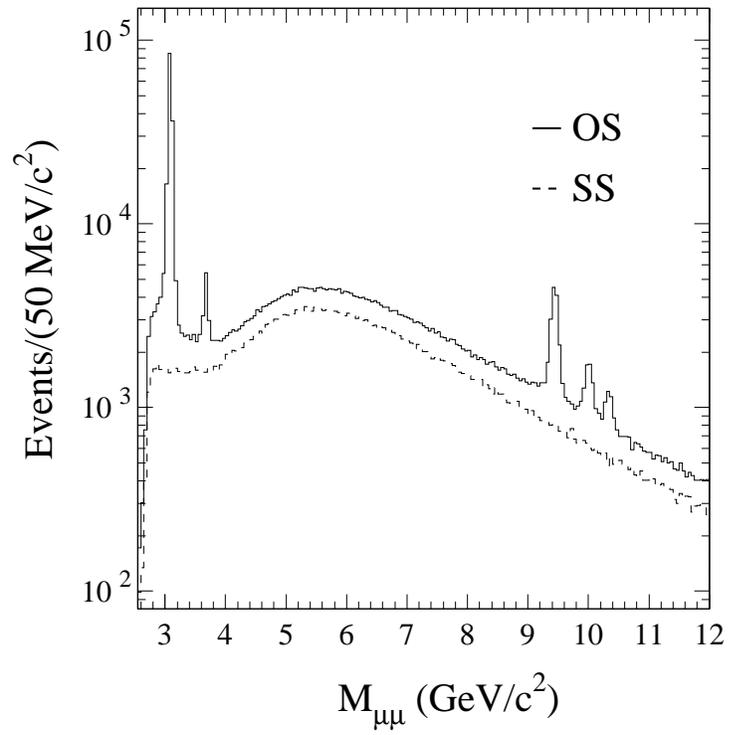}
 \caption[]{Invariant mass distribution of all muon pairs. OS indicate dimuons
            with opposite sign charge used in this analysis. Dimuons with same
            sign charge (SS) are also shown.}
 \label{fig:fig_dimu_1}
 \end{center}
 \end{figure}

\begin{figure}
 \begin{center}
 \vspace{-1.0cm}
 \leavevmode
 \includegraphics{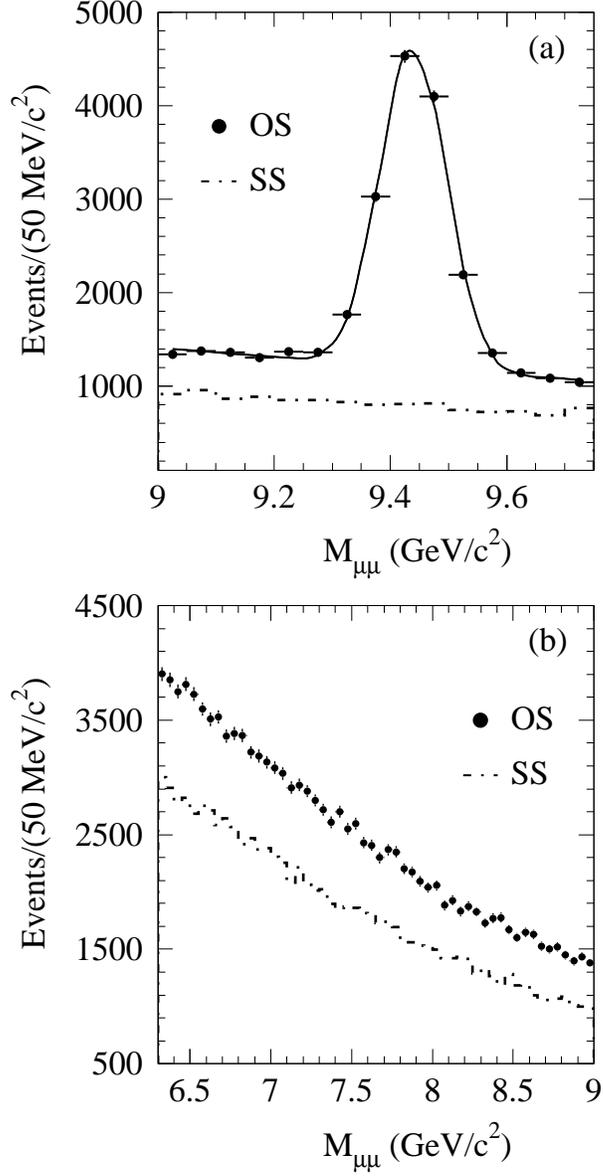}
 \caption[]{Invariant mass distributions of muon pairs at the $\Upsilon(1S)$
            (a) and in the region of interest for this study (b). The solid
            line represents the fit used to estimate the number of
            $\Upsilon(1S)$ mesons.}
 \label{fig:fig_dimu_2}
 \end{center}
 \end{figure}

  The dimuon invariant mass distribution is shown in Fig.~\ref{fig:fig_dimu_1}.
  The yield of $\psi$ mesons is much suppressed with respect to that of  
  $\Upsilon$ mesons. Because of the muon selection criteria, at the
  $J/\psi$ mass the kinematic acceptance decreases rapidly with $J/\psi$
  decreasing transverse momentum and vanishes at $p_T \simeq 5\; \gevc$.
  In contrast, for muon pairs with invariant  masses larger than
  $6.3\; \gevcc$, the kinematic acceptance does not depend on the dimuon 
  transverse momentum. Therefore, we avoid the uncertainty of modeling an
  acceptance that depends on the production kinematics by limiting our study
  to the mass region above $6.3\; \gevcc$. The dimuon 
  invariant mass distribution at the $\Upsilon(1S)$ and in the
  region of interest for this study are shown in Fig.~\ref{fig:fig_dimu_2}. 
  The number of $\Upsilon(1S)$ mesons in the data is derived by fitting
  a first order polynomial and a Gaussian function to the invariant mass
  distribution in Fig.~\ref{fig:fig_dimu_2}(a) with a binned maximum 
  likelihood method. The best fit returns $9838\pm 141\; \Upsilon(1S)$ mesons
  over a background of $5769$ events in the region 
  $9.3\leq M_{\mu^{+}\mu^{-}} \leq 9.55\; \gevcc$.
  The fit also returns $M_{\Upsilon(1S)} = 9439 \pm 1$ MeV/c$^2$ and a mass
  resolution $\sigma_{M} = 57 \pm 1$ MeV/c$^2$. This mass resolution is
  well modeled by a simulation of the process 
  $p\bar{p}\rightarrow \Upsilon(1S) X $. The simulation event generator
  produces $\Upsilon(1S)$ mesons with the transverse momentum distribution of
  the data~\cite{upsi-sec} and a flat rapidity distribution for $|y| \leq 1$.
  The generated events are processed  with the CDF detector
  simulation~\footnote{
  The simulation includes correction factors for the efficiency of the
  three-level trigger system, the efficiency for reconstructing tracks in the
  CTC and the different muon systems, effects due to instantaneous luminosity
  and to internal radiation from muons. These correction factors are
  parametrizations based on the data~\cite{upsi-sec,tsigma}.}
  ({\sc qfl}) described in detail in Ref.\cite{tsigma}.
  Events are then required to pass the same selection and reconstruction
  criteria imposed on the data. This simulation predicts a mass 
  resolution of $40$ MeV/c$^2$ for {\Large $\varepsilon$} states with a mass
  around $7.5\; \gevcc$.
 \clearpage
 \section{Background reduction} \label{sec:ss-muan} 
\begin{figure}
 \begin{center}
 \includegraphics*[width=\textwidth]{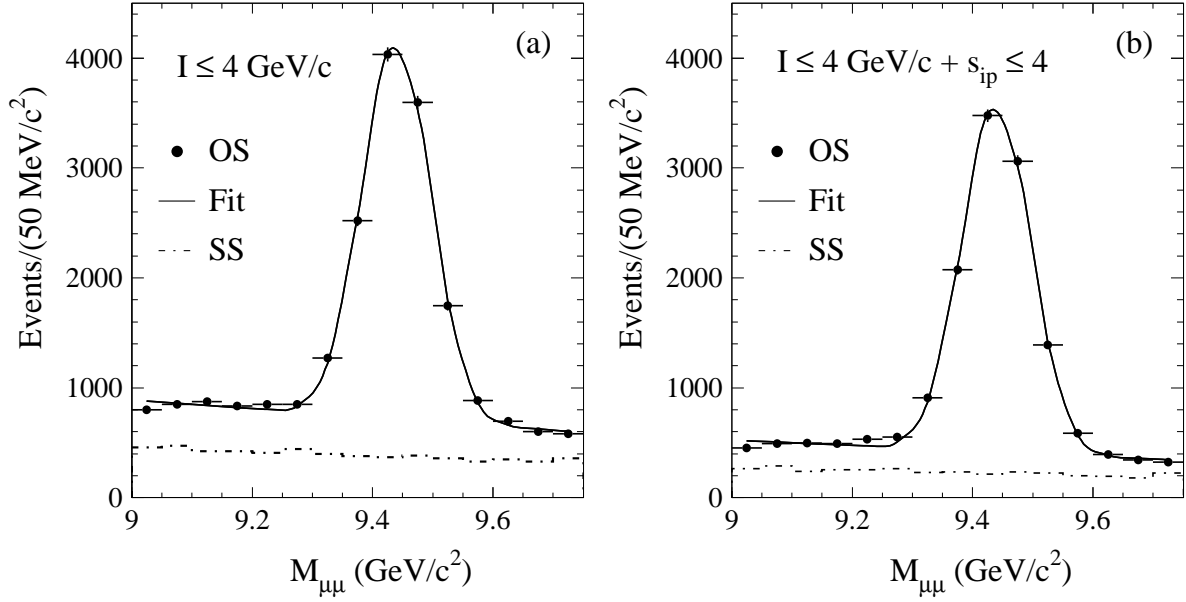}
 \caption[]{Invariant mass distributions of opposite ($\bullet$) and same sign
            charge (dot-dashed) muon pairs in the $\Upsilon~(1S)$ mass range
            after the isolation (a) and impact parameter cuts (b). The solid
            line is a fit using a Gaussian plus a first order polynomial 
            functions.}
 \label{fig:fig_dimu_3}
 \end{center}
 \end{figure}

   As outlined in Sec.~\ref{sec:onia}, the hypothetical signal of an
   {\Large $\varepsilon$} resonance of mass $\simeq 7.5\; \gevcc$ is expected
   to be at least $2$\% of the $\Upsilon(1S)$ yield, i.e. about $200$ events.
   Given the detector invariant mass resolution, the $200$ events have to be 
   integrated in a region of $150$ MeV/c$^2$ ($\pm 2 \sigma_{M}$). As shown 
   in Fig.~\ref{fig:fig_dimu_2}(b), three $50$ MeV/c$^2$ bins centered around 
   this mass contain approximately $8000$ events. This background can be
   largely suppressed because it is mostly contributed by $b\bar{b}$ and
   $c\bar{c}$ production. The measurement of the time-integrated 
   $B^{0}-\bar{B}^0$ mixing probability, reported in Ref.~\cite{bmix},
   also makes use of this data sample. From that study we estimate that 
   approximately $75$\% of the muon pairs arise from heavy flavor production.
   We use two intuitive criteria to reject dimuons arising from the decay of
   hadrons with  heavy flavor:
 \begin{enumerate}
 \item An isolation requirement. The isolation, $I$, is defined as the scalar
       sum of the transverse momenta of all the tracks in a cone of radius
       $R = \sqrt{\delta\phi^2 + \delta\eta^2}=0.4$ around the muon direction.
       We require that both muons have isolation $I \leq 4\; \gevc$.
 \item A promptness requirement. In contrast to $b$ and $c$-hadrons, the
       $\Upsilon(1S)$ and {\Large $\varepsilon$} mesons have negligible
       lifetime.  We select prompt muons by requiring the sum of the impact
       parameter significance of both muons, $s_{ip}$, to be less than
       4~\footnote{
       The track impact parameter $d$ is the distance of closest approach to
       the event primary vertex in the plane transverse to the beam line. The
       significance is defined as $d/\sigma_{d}$. The event primary vertex is
       determined as in the study in Ref.~\cite{bmix}.}.
       The impact parameter significance is estimated for muons with tracks
       reconstructed in the microvertex silicon detector (SVX),
       otherwise is set to zero in order not to lose events.
 \end{enumerate}
   The values of these cuts have been determined at the $\Upsilon(1S)$ mass
   (see Fig.~\ref{fig:fig_dimu_3}). As shown in Table~\ref{tab:tab_dimu_1},
   these cuts reduce the background  by more than a factor of three while
   retaining more than $90$\% of the $\Upsilon(1S)$ signal.
 \clearpage
  \begin{table}
 \begin{center}
 \caption[]{Numbers of $\Upsilon(1S)$ mesons and underlying-background events
            for different analysis cuts. Rates are evaluated by fitting the
            data with a Gaussian plus a first order polynomial function.
            The background is integrated over the mass region between $9.3$ 
            and $9.55\; \gevcc$. The last cut, $QC$, is used in the
            {\Large $\varepsilon$} search. Efficiencies are calculated with
            respect to the number of $\Upsilon(1S)$ candidates in the first
            row.}
 \begin{ruledtabular}
 \begin{tabular}{lccc}
 Cut             & $\Upsilon(1S)$ candidates & Background & Efficiency(\%)  \\
 None            &     $9838 \pm 141$        & $5769$     &         \\
 $I$             &     $9821 \pm 129$        & $3345$     &  $99.8$ \\
 $QC=I+ s_{ip}$  &     $9064 \pm 118$        & $1842$     &  $92.1$ \\
 \end{tabular}
 \end{ruledtabular}
 \label{tab:tab_dimu_1}
 \end{center}
 \end{table}

\clearpage
  \section{Estimate of the upper limit on 
           $\Gamma_l^{\mbox{\Large $\varepsilon$}}$} \label{sec:ss-mueps} 
\begin{figure}
 \begin{center}
 \leavevmode 
 \includegraphics{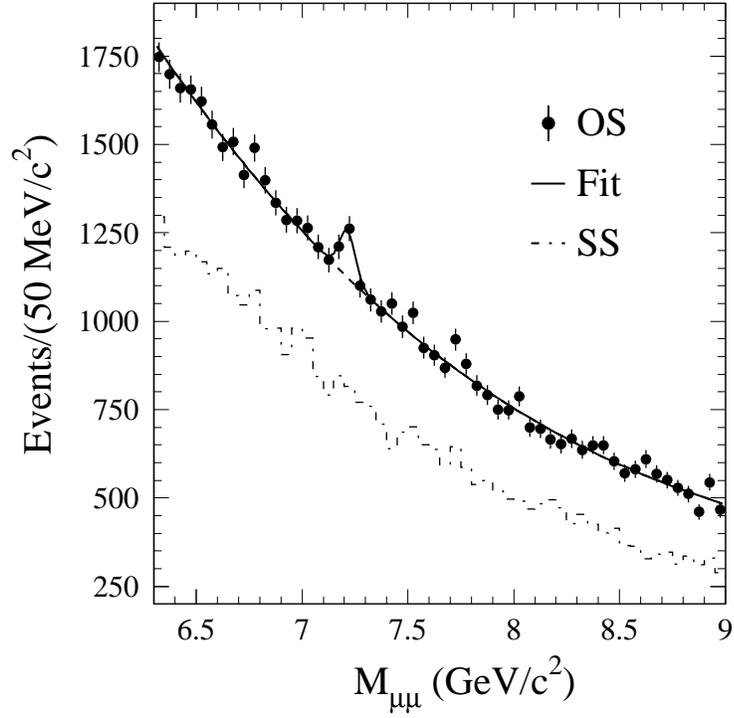}
 \caption[]{Invariant mass distribution of opposite ($\bullet$) and same sign
            charge (dot-dashed) muon pairs which pass the isolation and impact
            parameter cuts. The solid line represents the fit described
            in Sec.~\ref{sec:ss-mucheck}.}
 \label{fig:fig_dimu_4}
 \end{center}
 \end{figure}

   Figure~\ref{fig:fig_dimu_4} shows the invariant mass distribution of muon
   pairs in the region between $6.3$ and $9.0\; \gevcc$ after the isolation
   and impact parameter significance cuts.

   We use a binned maximum likelihood method to fit the mass spectrum in
   Fig.~\ref{fig:fig_dimu_4} with a fourth order polynomial, which serves the
   purpose of modeling a smooth background, plus a Gaussian function, which
   searches for narrow resonances. We perform $54$ fits, in which we constrain
   the Gaussian peak to the center of each of the $54$ mass bins of
   Fig.~\ref{fig:fig_dimu_4}; in each fit, we force the Gaussian width to the
   simulated resolution of the detector for that mass. For each mass bin, we
   use the integral of the Gaussian function and its error returned by the 
   best fit to derive $N_{ul}$, the $90$\% credibility upper limit to the 
   number of events contributed by a narrow resonance centered in that mass
   bin~\footnote{ 
   The integral of the fit likelihood from $N_{ul}$ to infinity is $10$\% of 
   the integral of the fit likelihood from $0$ to infinity. The justification
   for this procedure is Bayesian with a prior that is zero for negative 
   resonance cross sections and flat for positive ones.}.
   We evaluate the ratio of the geometric and kinematic acceptance for an 
   {\mbox{\Large $\varepsilon$}} resonance to that for the $\Upsilon(1S)$
   meson with the simulation described at the end of Sec.~\ref{sec:ss-mudat}. 
   In the event generator, an unpolarized resonance is produced with a flat
   rapidity distribution ($|y| \leq 1$) and a transverse momentum distribution
   that is rescaled from that of the $\Upsilon(1S)$ data so that
   $<p_T^{\mbox{\Large $\varepsilon$}}>/<p_T^{\Upsilon(1S)}>=
   m_{\mbox{\Large $\varepsilon$}}/m_{\Upsilon(1S)}$. 
   For $J/\psi$ mesons, this rescaling procedure predicts a $d\sigma/dp_T$
   distribution that decreases more rapidly with increasing momenta than the
   distribution of the data. However, a poor modeling of the transverse
   momentum distribution is not a cause of error because the kinematic
   acceptance does not depend on the  {\mbox{\Large $\varepsilon$}}
   transverse momentum. As a cross-check, we also generated 
   {\mbox{\Large $\varepsilon$}} resonances using the shape of the transverse
   momentum distribution of the $\Upsilon(1S)$ data and verified that they 
   return consistent acceptance values. The geometric and kinematic acceptance
   increases from 8.3\% at 6.3 $\gevcc$ to 11.2\% at 9.0 $\gevcc$ (it is
   11.7\%  at $\Upsilon(1S)$ mass). The parametrized correction factors for
   the trigger and reconstruction efficiencies, discussed at the end of 
   Sec.~\ref{sec:ss-mudat}, depend little on the muon $p_T$. Including these
   effects in the simulation, the acceptance is 4.7\% at 6.3 $\gevcc$ and 
   6.3\% at 9.0 $\gevcc$ (it is 6.6\% at $\Upsilon(1S)$ mass).
   The parametrized correction factors to the acceptance have approximately
   a 6\% uncertainty. However, in the mass range considered in this study, 
   the ratio of the simulated {\mbox{\Large $\varepsilon$}} to  $\Upsilon(1S)$
   acceptances is not affected by this uncertainty.
 
\begin{figure}
 \begin{center}
 \leavevmode
 \includegraphics{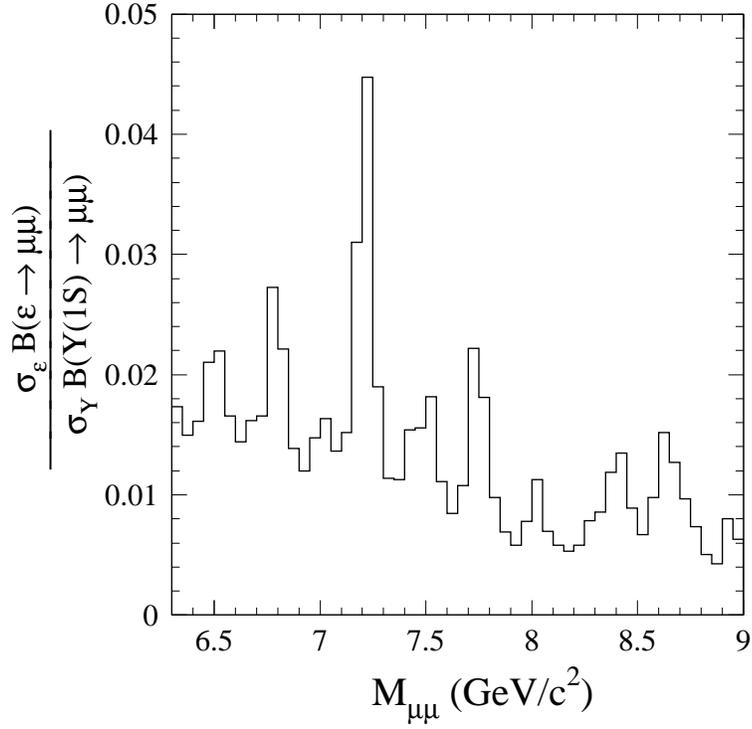}
 \caption[]{Bayesian $90$\% upper limit to 
            $\frac { \sigma_{\mbox{\Large $\varepsilon$}}
            \; B(\mbox{{\Large $\varepsilon$}} \rightarrow \mu\mu)}
            {\sigma_{\Upsilon(1S)}\;  B(\Upsilon(1S) \rightarrow \mu \mu)}$
            as a function of the {\Large $\varepsilon$} mass.}
 \label{fig:fig_lim1}
 \end{center}
 \end{figure}

\begin{figure}
 \begin{center}
 \leavevmode
 \includegraphics{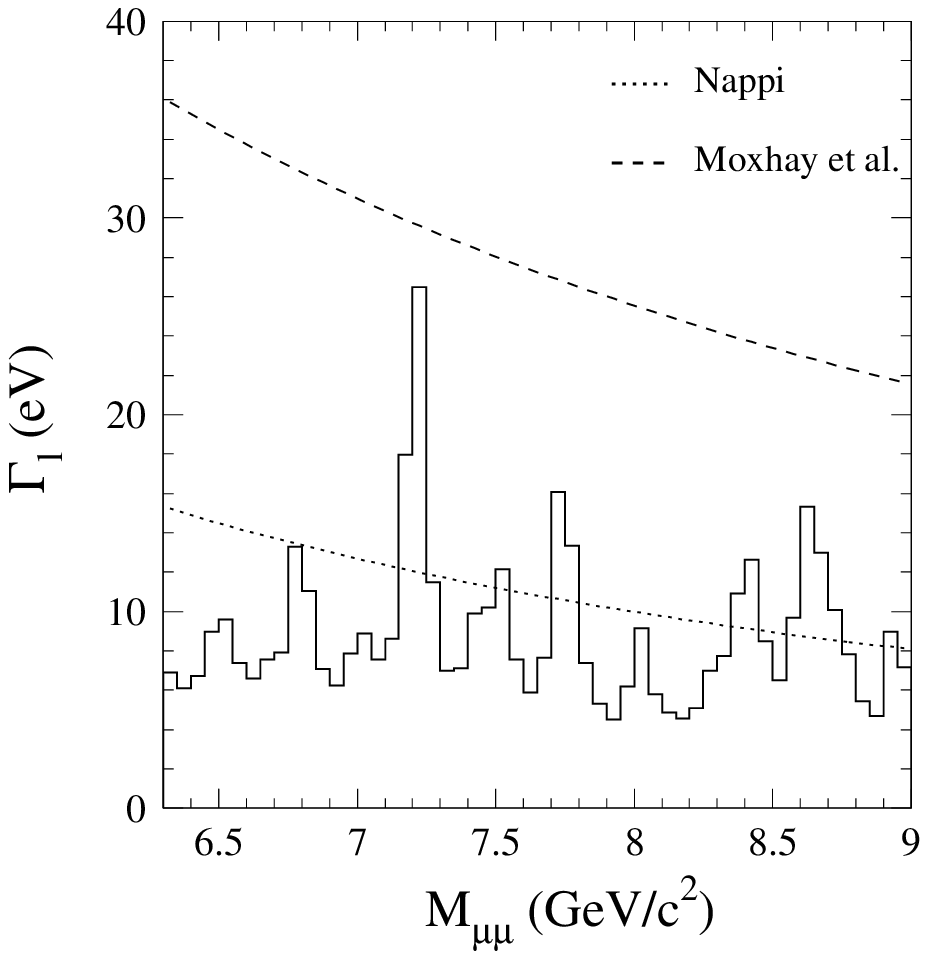}
 \caption[]{Bayesian $90$\% upper limit to 
            $\Gamma_l^{\mbox{\Large $\varepsilon$}}$ (histogram). The dashed
            and dotted lines represent the leptonic widths of 1$^{--}$ bound
            states of scalar quark predicted in Refs.~\cite{nappi} 
            and~\cite{moxhay}, respectively.}
 \label{fig:fig_lim2}
 \end{center}
 \end{figure}

\begin{figure}
 \begin{center}
 \leavevmode
 \includegraphics{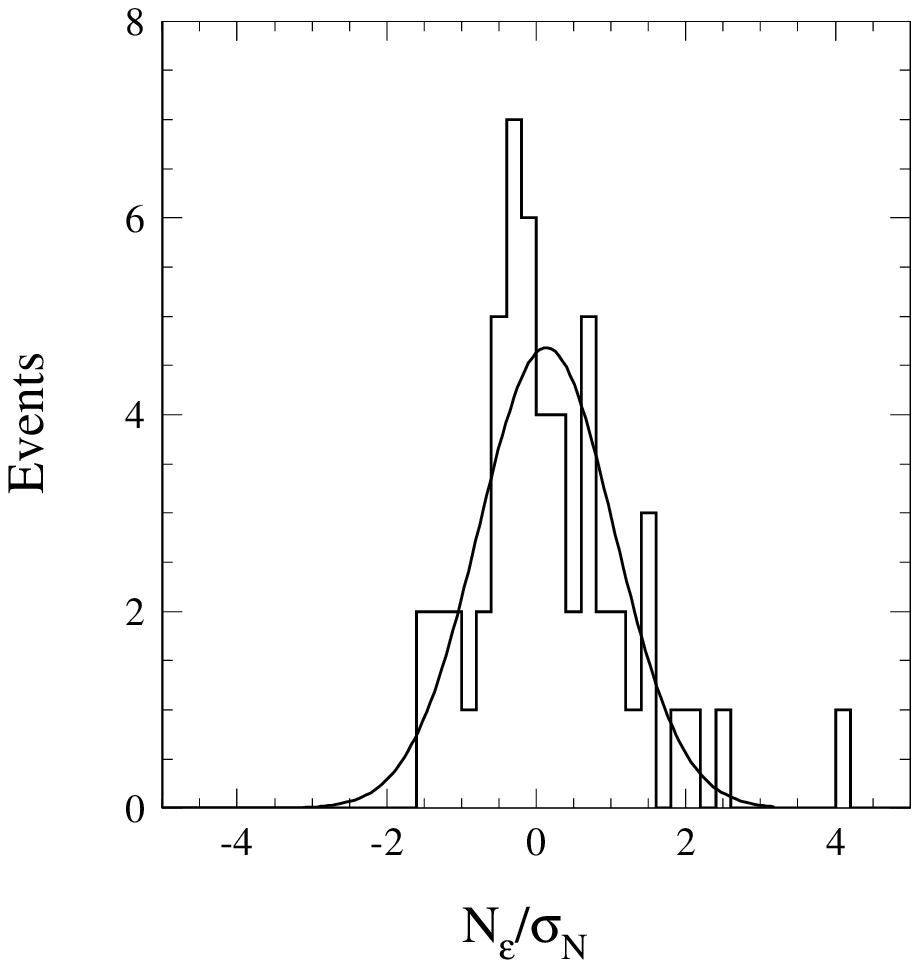}
 \caption[]{Distribution of $N_{\mbox{\Large $\varepsilon$}}$/$\sigma_N$ (see
            text). The solid line is a Gaussian function with unit width.}
 \label{fig:fig_lim3}
 \end{center}
 \end{figure}

  The ratio of $N_{ul}$ to the number of observed $\Upsilon(1S)$ mesons,
  corrected for the relative acceptance, provides the $90$\% credibility upper
  limit to
       $\sigma_{\mbox{\Large $\varepsilon$}}
        \; B(\mbox{{\Large $\varepsilon$}} \rightarrow \mu\mu)/ 
        \sigma_{\Upsilon(1S)}\;  B(\Upsilon(1S) \rightarrow \mu \mu)$ 
  shown in Fig.~\ref{fig:fig_lim1}.
  Figure~\ref{fig:fig_lim2} shows the $90$\% credibility upper limit to
  $\Gamma_l^{\mbox{\Large $\varepsilon$}}$ derived using equation (1).
 We note that equation (1) tends to underestimate the  {\Large $\varepsilon$}
 production cross section. One can verify this by predicting the $J/\psi$
 production cross section from that of the $\Upsilon(1S)$ meson that is
 about two order of magnitude smaller. The  $\Upsilon(1S)$ integrated cross
 section for $|y| \leq 0.6$ is measured to be $34.6\pm 2.5$ nb~\footnote{
 Average of the measurements in Refs.~\cite{upsi-sec}.}.
 The analogous cross section for the  $J/\psi$ meson is 
 $3060\pm 460$ nb~\footnote{
 We use the result reported in  Ref.~\cite{psi-sec}.
 The measurement is performed at $\sqrt{s}=1.96$
 TeV, and we rescale the published value by the 10\% expected increase of the 
 cross section from 1.8 to 1.96 TeV~\cite{psi-sec}. We also extrapolate the result 
 to $p_T^{\rm min}=0\; \gevc$ assuming that the fraction of prompt $J/\psi$
 remains constant for transverse momenta smaller than $1.5 \;\gevc$.},
 whereas equation (1) predicts $1630 \pm 116$ nb.
 If the {\Large $\varepsilon$} production cross sections predicted by eqn. (1)
 were also a factor of two smaller than the data, the 
 $\Gamma_l^{\mbox{\Large $\varepsilon$}}$ limits set by our study would be
 a factor of two smaller than those indicated in Fig.~\ref{fig:fig_lim2}.  
   Figure~\ref{fig:fig_lim3} shows the distribution of 
   $N_{\mbox{\Large $\varepsilon$}}$, the number of events attributed to a
   narrow resonance, divided by the error $\sigma_N$ returned by the best fit
   for the $54$ considered mass bins. With the exception of a point at
   $4.1\; \sigma$, this distribution is consistent with a Gaussian function
   of unit width. Therefore,  it seems fair to assume that  the distribution
   of the $90$\% upper limits in Fig.~\ref{fig:fig_lim2} is statistically 
   consistent with the average upper limit 
   $\Gamma_l^{\mbox{\Large $\varepsilon$}}= 8$ eV that corresponds to those
   cases with $N_{\mbox{\Large $\varepsilon$}}/\sigma_N =0$ in
   Fig.~\ref{fig:fig_lim3}. The  $4.1\; \sigma$ fluctuation occurs at the 
   mass of $7.25\; \gevcc$.
\clearpage
 \section{Study of the $7.25\; \gevcc$ fluctuation}\label{sec:ss-mucheck}
  In this section we explore the possibility that the anomalously large upper 
  limit at $7.25\; \gevcc$ is due to a real signal. A fit, which uses a 
  Gaussian function with a fixed $38$ MeV/c$^2$  resolution, returns 
  $M_{\mbox{\Large $\varepsilon$} } = 7.22 \pm 0.01\; \gevcc$ and a signal 
  of $250\pm 61$ events over a smooth background of $3355$ events,
  extrapolated in the $150$ MeV/c$^2$ region between $7.15$ and $7.3\; \gevcc$
  (see Fig.~\ref{fig:fig_dimu_4}). The probability that $3355$ background
  events fluctuate to no less than $3605$ is $8\times 10^{-6}$
  ($4.3\; \sigma$). Since the mass range examined in 
  Fig.~\ref{fig:fig_dimu_4} includes $52$ almost independent combinations of
  three consecutive $50$ MeV/c$^2$ bins, the probability of obtaining an equal
  or larger statistical fluctuation in the inspected mass window is
  approximately $4.1 \times 10^{-4}$ ($3.5\; \sigma$).

  According to the simulation the acceptance for a $7.2\ \gevcc$ resonance 
  relative to that for the $\Upsilon(1S)$ meson is 
    $A_{\mbox{\Large $\varepsilon$}} = 0.78 \times A_{\Upsilon(1S)}$. 
  It follows that
  \[  \sigma_{\mbox{\Large $\varepsilon$}} \;
        B(\mbox{\Large $\varepsilon$} \rightarrow \mu\mu) =
        (3.6 \pm 0.9) \times 10^{-2} \times \sigma_{\Upsilon(1S)}\;
        B(\Upsilon(1S) \rightarrow \mu \mu)
  \]
  We note that this value is in agreement with the theoretical expectation for
  a bound state of charge$-1/3$ squarks.

\begin{figure}
 \begin{center}
 \leavevmode
 \vspace{-1.0cm}
 \includegraphics*[width=\textwidth]{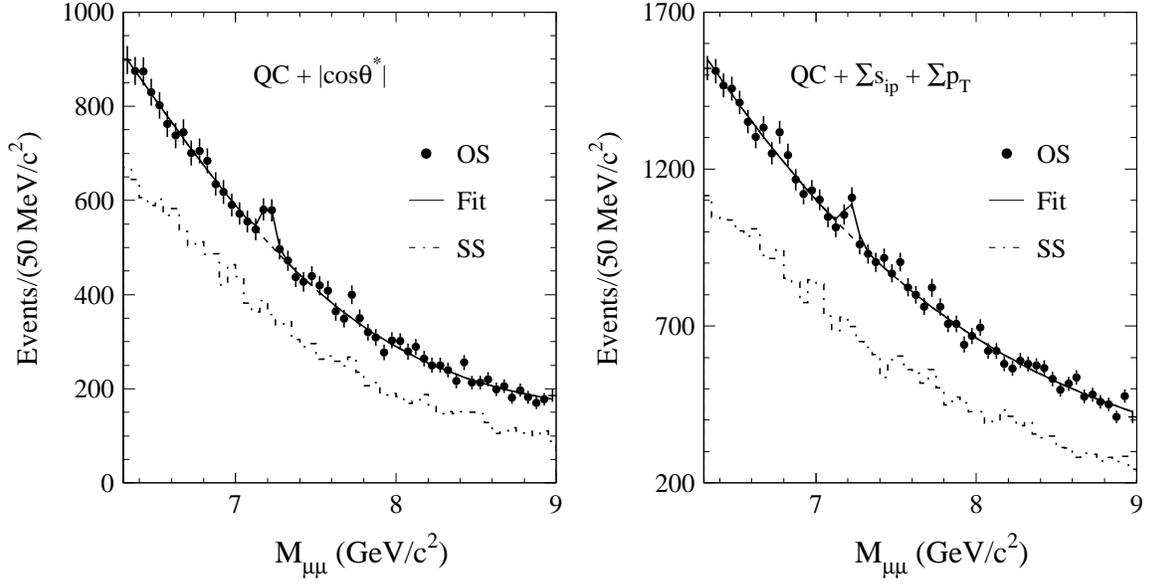}
 \vspace{-1.5cm}
 \caption[]{Invariant mass distributions of muon pairs after applying the
            first and second cuts described in Table~\ref{tab:tab_dimu_2}.
            The solid line is a fit to the data described in the text. }
 \label{fig:fig_dimu_5_a}
 \end{center}
 \end{figure}

\begin{figure}
 \begin{center}
 \vspace{-1.5cm}
 \leavevmode
 \includegraphics{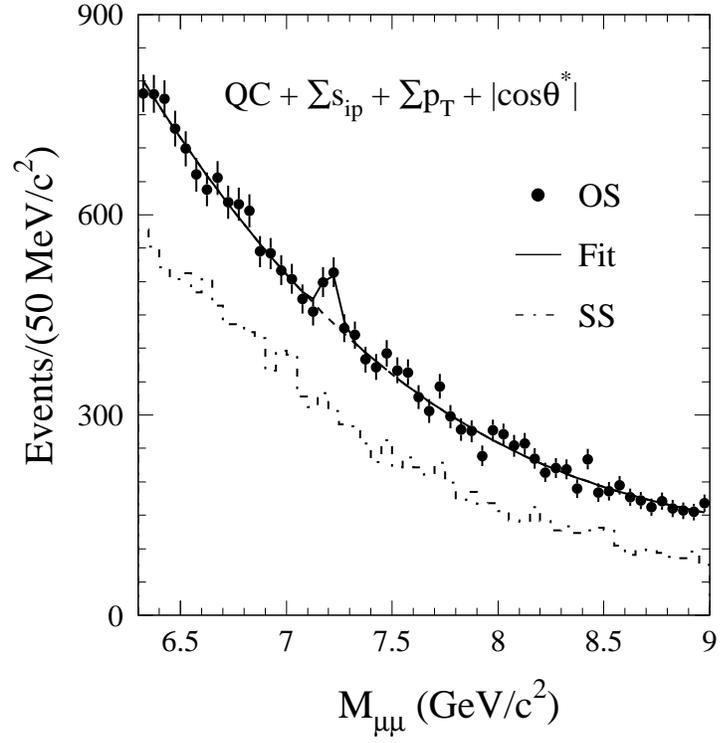}
 \vspace{-0.5cm}
 \caption[]{Invariant mass distributions of opposite sign charge muon pairs
            after applying the third cut described in 
            Table~\ref{tab:tab_dimu_2}.}
 \label{fig:fig_dimu_5_b}
 \end{center}
 \end{figure}

  We have investigated three additional selection cuts that reduce the number
  of events by more than a factor of two and compare the effect of these cuts
  on the number of {\Large  $\varepsilon$} and $\Upsilon(1S)$ candidates:
 \begin{enumerate}
 \item $|\cos(\theta^{\ast})| \leq$ 0.4, where $\theta^{\ast}$ is the polar
       angle between the $\mu^{+}$ and {\Large  $\varepsilon$} directions in
       the {\Large  $\varepsilon$} center-of-mass system. This cut reduces the
       number of events by simply selecting a particular sector of the phase
       space.
 \item ${\displaystyle \sum_i} p_T \leq 40\; \gevc$ and
       ${\displaystyle \sum_j} s_{ip}^j \leq$ 30, where
       ${\displaystyle \sum_i} p_T^i$ is the scalar sum of the transverse
       momentum of all tracks $i$ originating from the same vertex as the
       muon pair and ${\displaystyle \sum_j} s_{ip}^j$ is the sum
       of the impact parameter significance of all tracks $j$ not used to
       define the primary vertex of the event~\cite{vertex}. This cut is
       intended to further suppress the heavy flavor background by rejecting 
       events in which the muon pair carries a small fraction of the total
       transverse momentum or is produced in association with additional
       long-lived particles.
 \item cut \#1 + cut \#2
 \end{enumerate}
   The effect of these cuts is shown in Figs.~\ref{fig:fig_dimu_5_a}
   and~\ref{fig:fig_dimu_5_b}, and is compared to the result for the
   $\Upsilon(1S)$ meson in Table~\ref{tab:tab_dimu_2}.
 \clearpage
  \begin{table}
 \begin{center}
 \caption[]{Yield of {\Large $\varepsilon$} and $\Upsilon(1S)$ candidates
            for different analysis cuts. The underlying backgrounds are 
            fitted with polynomial functions and integrated over the mass
            ranges $7.15 - 7.3$ and $9.3-9.55\; \gevcc$.}
 \begin{ruledtabular}
 \begin{tabular}{lcccc}
  Cut  & {\Large $\varepsilon$} & Background & $\Upsilon(1S)$  & Background\\
 $QC$  &  $249.7 \pm 60.9$      & $3355.0$   & $9064 \pm 118$  & $1842$  \\
 \# 1  &  $160.5 \pm 41.8$      & $1508.0$   & $3910 \pm  90$  & $ 611$  \\
 \# 2  &  $206.2 \pm 57.0$      & $2948.0$   & $8667 \pm 136$  & $1587$  \\
 \# 3  &  $144.5 \pm 39.0$      & $1311.0$   & $3699 \pm  87$  & $ 561$  \\
 \end{tabular}
 \end{ruledtabular}
 \label{tab:tab_dimu_2}
 \end{center}
 \end{table}

 \clearpage
 \section{Conclusions} \label{sec:concl}
   We have investigated the invariant mass spectrum of dimuons collected
   by the CDF experiment at the Tevatron collider to improve the limit to the
   existence of narrow resonances set by the experiments at the SPEAR 
   $e^{+} e^{-}$ collider. In the mass range $6.3-9.0 \; \gevcc$, we derive
   90\% upper credible limits to the ratio of the production cross section
   times muonic branching fraction of possible narrow resonances to that of 
   the $\Upsilon(1S)$ meson. In this mass range, the average limit varies
   from 1.7 to 0.5\%. Assuming that 
      $\sigma_{\mbox{\Large $\varepsilon$}}=
      \sigma_{\Upsilon(1S)} \times (m_{\Upsilon(1S)}/
       m_{\mbox{\Large $\varepsilon$}})^3 
      \times \Gamma_{\mu }^{\mbox{\Large $\varepsilon$}}/ 
      \Gamma_{\mu }^{\Upsilon(1S)}$,
   these limits correspond to an average $90$\% upper credible limit of $8$ eV 
   to the leptonic width of possible resonances. An exception is the mass
   region around $7.2\; \gevcc$ where we observe a bump of $250 \pm 61$ events
   with a width consistent with the detector resolution. The size of the 
   excess is consistent with the theoretical expectation for the production
   of a $1^{--}$ p-wave resonance but its statistical significance 
   ($3.5\; \sigma$) is not sufficient to claim the discovery of a new particle.
 \section{Acknowledgments}
  We thank the Fermilab staff, the CDF collaboration, and their technical
  staff for their contributions. This work was supported by the
  U.S.~Department of Energy and National Science Foundation; the Istituto
  Nazionale di Fisica Nucleare; and the Ministry of Education, Culture,
  Sports, Science and Technology of Japan.



\begin{thebibliography}{99}
\bibitem{vertex}  Tracks with impact parameter significance smaller than
                   three are used to reconstruct the primary vertex; we reject
                   tracks with impact parameter larger than $0.15$ cm.
\end{thebibliography}
 \end{document}